\documentclass[draft]{agujournal2019}
\usepackage{url} %this package should fix any errors with URLs in refs.
\usepackage{lineno}
\usepackage[inline]{trackchanges} %for better track changes. finalnew option will compile document with changes incorporated.
\usepackage{soul}
\usepackage{xcolor}
\usepackage[square,sort]{natbib}
\usepackage{fancyhdr}

\fancypagestyle{acceptedmanuscript}{%
  \fancyhf{}%
  \fancyhead[C]{\small\itshape Accepted Manuscript}%
  \fancyfoot[C]{\thepage}%
}
\fancypagestyle{plain}{%
  \fancyhf{}%
  \fancyhead[C]{\small\itshape Accepted Manuscript}%
  \fancyfoot[C]{\thepage}%
}
\draftfalse

\journalname{JGR: Space Physics}

\begin{document}

% This file contains the accepted, unformatted manuscript version.
% It is not the Version of Record published by AGU/Wiley.

%%%%%%%%%%%%%%%%%%%%%%%%%%%%%%%%%%%%%%%%%%%%%%%
%  TITLE
%
% (A title should be specific, informative, and brief. Use
% abbreviations only if they are defined in the abstract. Titles that
% start with general keywords then specific terms are optimized in
% searches)
%
%%%%%%%%%%%%%%%%%%%%%%%%%%%%%%%%%%%%%%%%%%%%%%%

% Example: \title{This is a test title}

\title{Modelling of the variability of D-region ionospheric electron density during solar cycle-24}

%%%%%%%%%%%%%%%%%%%%%%%%%%%%%%%%%%%%%%%%%%%%%%%
%
%  AUTHORS AND AFFILIATIONS
%
%%%%%%%%%%%%%%%%%%%%%%%%%%%%%%%%%%%%%%%%%%%%%%%

% Authors are individuals who have significantly contributed to the
% research and preparation of the article. Group authors are allowed, if
% each author in the group is separately identified in an appendix.)

% List authors by first name or initial followed by last name and
% separated by commas. Use \affil{} to number affiliations, and
% \thanks{} for author notes.
% Additional author notes should be indicated with \thanks{} (for
% example, for current addresses).

% Example: \authors{A. B. Author\affil{1}\thanks{Current address, Antartica}, B. C. Author\affil{2,3}, and D. E.
% Author\affil{3,4}\thanks{Also funded by Monsanto.}}

\authors{Sayak Chakraborty\affil{1}, Sourav Palit\affil{1}, Semontee Deb\affil{2}, Tamal Basak\affil{1} }

% \affiliation{1}{First Affiliation}
% \affiliation{2}{Second Affiliation}
% \affiliation{3}{Third Affiliation}
% \affiliation{4}{Fourth Affiliation}

\affiliation{1}{Indian Centre for Space Physics, 466 Barakhola, Netai Nagar, Kolkata 700099, India}
\affiliation{2}{Department of Atmospheric Sciences, University of Calcutta, Kolkata 700019, India}
%(repeat as many times as is necessary)

% Corresponding author mailing address and e-mail address:

% (include name and email addresses of the corresponding author.  More
% than one corresponding author is allowed in this LaTeX file and for
% publication; but only one corresponding author is allowed in our
% editorial system.)

% Example: \correspondingauthor{First and Last Name}{email@address.edu}

\correspondingauthor{Tamal Basak}{tamalbasak@gmail.com}

\justifying
%%%%%%%%%%%%%%%%%%%%%%%%%%%%%%%%%%%%%%%%%%%%%%%
% KEY POINTS
%%%%%%%%%%%%%%%%%%%%%%%%%%%%%%%%%%%%%%%%%%%%%%%
%  List up to three key points (at least one is required)
%  Key Points summarize the main points and conclusions of the article
%  Each must be 140 characters or fewer with no special characters or punctuation and must be complete sentences

% Example:
% \begin{keypoints}
% \item	List up to three key points (at least one is required)
% \item	Key Points summarize the main points and conclusions of the article
% \item	Each must be 140 characters or fewer with no special characters or punctuation and must be complete sentences
% \end{keypoints}

\begin{keypoints}
\item Variability of solar Lyman-$\alpha$ and X-ray due to solar cycle 24
\item D-region ionization in presence of solar Lyman-$\alpha$ and X-ray
\item Solar cycle 24 induced D-region electron density variation
\end{keypoints}

%%%%%%%%%%%%%%%%%%%%%%%%%%%%%%%%%%%%%%%%%%%%%%%
%
%  ABSTRACT and PLAIN LANGUAGE SUMMARY
%
% A good Abstract will begin with a short description of the problem
% being addressed, briefly describe the new data or analyses, then
% briefly states the main conclusion(s) and how they are supported and
% uncertainties.

% The Plain Language Summary should be written for a broad audience,
% including journalists and the science-interested public, that will not have 
% a background in your field.
%
% A Plain Language Summary is required in GRL, JGR: Planets, JGR: Biogeosciences,
% JGR: Oceans, G-Cubed, Reviews of Geophysics, and JAMES.
% see http://sharingscience.agu.org/creating-plain-language-summary/)
%
%%%%%%%%%%%%%%%%%%%%%%%%%%%%%%%%%%%%%%%%%%%%%%%

%% \begin{abstract} starts the second page

\begin{abstract}
Solar cycle variation of earth's atmosphere, particularly the ionosphere is of particular interest in the field of space science and space weather studies. In this article, we present the outcome of our detailed quantitative study on solar cycle variation of lower ionospheric properties using numerical investigation. First, we seek to model and compare the collective D-region ionization rates ($q$'s) due to the individual contributions from the ionizing sources, namely, (i) solar extreme ultraviolet (EUV) radiation including the Lyman-$\alpha$ irradiation and (ii) solar X-ray irradiation throughout the $24^{th}$ solar cycle (C24). Then, we compute the electron density ($N_e$) profiles using the ionization rates. We report significant solar cycle variation in ionization rate and electron density profiles for the entire span of C24. We use the sunspot number (SPN) profile to substantiate the finer details of $N_e$ profile during C24. For the segments of the D-region above two different geographic coordinates, we report that $N_e$ profiles show a consistent `dual peak' nature. It is very similar to the SPN profile during C24. As a next-order validation, we compare our modelled $N_e$ profiles with their IRI-2020 counterpart ($N_{e,iri}$s). Their overall trends are found to be in agreement. Finally, we discuss the response of the D-region in terms of $N_e$ due to C24. The work lays the foundation for our upcoming studies on D-region response to solar cycle variation with Very Low Frequency (VLF) observation.
\end{abstract}

\section*{Plain Language Summary}
The solar cycle is a periodic change in solar activity with a periodicity of about 11 years. An observable indicator of the solar cycle is the temporal variation in Sunspot Number (SPN). During the supposed maxima period of most of the observed cycles, two peaks have been detected in the sunspot data. The gap between these two peaks is named as {\it{Gnevyshev gaps}}. The D-region is the lowermost part of the ionosphere with the altitude distribution from, roughly, $60$ to $90$ km. We develop a numerical model to study the long-term variability of the D-region ionosphere during solar cycle 24 (C24). The D-region is maximally ionised by solar extreme ultraviolet (EUV) and X-ray radiation. We estimate the net ionization rate ($q$) due to both of these two types of solar radiation in this region. We model the electron density ($N_e$) profile and check the long-term effects due to C24. We found, just like the `dual peak' in SPN, $N_e$ profiles also show similar `dual peak' consistently across different parts of the D-region above different geographic coordinates. We also investigate the altitude ($h$) dependency of ionization rate ($q$) and $N_e$. We infer that the D-region ionosphere varies appositely with solar activity phases during C24.

%%%%%%%%%%%%%%%%%%%%%%%%%%%%%%%%%%%%%%%%%%%%%%%
%
%  BODY TEXT
%
%%%%%%%%%%%%%%%%%%%%%%%%%%%%%%%%%%%%%%%%%%%%%%%

\section{Introduction}
 The sun's core is a site of spontaneous nuclear fusion that releases a tremendous amount of energy in the form of heat and light. Owing to the complexity of its internal structure and magnetic field configurations, the sun manifests a variety of dynamic phenomena, collectively known as solar activity. This includes the appearance of sunspots, which are comparatively cooler regions on the solar surface with intense magnetic activity and flux. It appears darker than the neighboring regions on the solar surface. Some highly energetic eruptions of radiations and plasma in the form of solar flares and Coronal Mass Ejections (CMEs) occur at and around those spots. The occurrence of such activities, as well as, the overall solar radiation level go through a periodic variation with a periodicity of about 11 years, termed as `solar cycle'. Each solar cycle goes through a complete phase of high and low activities. The higher activity is found during the middle of the cycle. The sunspot number (SPN) is considered as a prominent and observable indicator of the solar cycles \citep[etc.]{newkirk82, bray64, wittmann87, ribes93, solanki03}. 
 
 As the primary source of energy governing the earth's atmospheric dynamics and evolution is the sun, any long-term variation in the solar activity level, like the solar cycle, significantly impacts the earth's atmosphere. Particularly, the ionosphere, comprising partially ionized regions within the atmosphere, is anticipated to experience the most notable effects from the fluctuations in solar activity and radiation levels across the cycle.

 The D-region of the lower ionosphere is mostly consists of molecular oxygen, molecular nitrogen, nitric oxide, and other compounds having different ionization cross-sections. Among them, nitric oxide is one of the principal constituents in the upper part of the D-region. It has an ionization potential of $9.25$ eV. So, the Lyman-$\alpha$ radiation is primarily responsible for ionizing the upper part of the D-region. The penetration and hence, the ionization by solar radiation at lower altitudes ($\sim 75$ km) is mostly possible for solar radiation of wavelengths between $110$-$121.6$ nm (\cite{nicolet60}). The absorption cross-section of X-ray having wavelengths shorter than $1$ nm is generally less than $10^{-23}$ cm$^2$ at $1$ nm to $10^{-26}$ cm$^2$ at $0.1$ nm (\cite{nicolet60}). Therefore, this spectral range can penetrate through the entire altitude range of the D-region ionosphere and cause ionization.  
 
 During periods of low solar activity, X-ray energies are insufficient to fully penetrate the D-region. Then, the solar UV radiation, particularly Lyman-$\alpha$, remains a primary ionizing agent in the D-region. However, as solar activity grows, the X-ray energy level increases and begins to play a significant role in ionization. Consequently, both UV and X-rays are essential for ionizing the D-region. Therefore, when we model the D-region variability during a solar cycle, we must take into account the contributions in ionospheric ionization due to both UV and X-rays.

 Importantly, during periods of higher activity, more than one (usually 2) maxima have been noted in SPN data. It led solar scientists to attribute this to the north-south hemispheric asymmetry in the sun \citep[etc.]{waldmeier57,waldmeier71,roy77}. A model of the complex magnetic field structure of the sun has also been proposed for explaining the sudden minimum taking place between two maxima. It is formally termed as {\it{Gnevyshev gaps}} \citep{gnevyshev63}. After looking at the activities at each of the solar hemispheres separately, it has been confirmed that the double maxima are not the result of the out-of-phase pursuit between the hemispheres. Actually, the gaps are present individually in both the hemispheres \citep[etc.]{norton10, ravindra21}.

 The effects of the 11-year cycle of solar activity on the earth's ionosphere have been studied in terms of Total Electron Content (TEC) (derived from the Global Positioning System, GPS), the International Reference Ionosphere (IRI-2020) model and the Thermosphere-Ionosphere Electrodynamics General Circulation Model (TIE-GCM) during C24 \citep{rao19}. They filtered out the effects of solar flares and geomagnetic storms from the corresponding TEC measurements for further analysis. They reported such dual peaks in the F10.7 cm index and TEC as obtained through the IRI-2020 model and their observed TEC values. The F10.7 index is a measure of the noise level generated by the sun at a wavelength of $10.7$ cm in the earth's orbit. It is an established parameter for measuring solar activity levels. They also reported a linear relationship between TECs and the F10.7 index. \citet{araujo11} investigated the solar minima in between the solar cycles $23$ and $24$ by studying two major ionospheric parameters, namely, vTEC and NmF2. They also analysed the $81$-day running average value of the F10.7 index with vTEC and reported similar dual peaks. They claimed that the ionospheric response is in accordance with the solar conditions. They reported the decrement of vTEC to be consistent, but the NmF2 was not as consistent as vTEC during the minima period. \citet{kumar18} investigated the response of the VLF signals for various solar flares during both low and high solar activity phases. They did it for signal propagation paths from NWC and NLK transmitters respectively to the receiving station at Fiji. They reported that the probability of occurring a solar flare is higher when SPN is relatively high. They also reported a notable solar activity dependency of the sub-ionospheric radio signals with greater VLF amplitude perturbations during solar flares in the lower solar activity phases than that of the higher. They attributed such dependency to the solar background radiation level. \citet{venka23} studied the effect of X-class solar flares (occurred between 2008 and 2016) on the D-region ionosphere. They used the VLF signal propagation from the NWC transmitter to the receiving station at Prayagraj, India path for this analysis. They reported multiple peaks in Lyman-$\alpha$ flux of $128$ days smoothed time variation. They claimed a logarithmic decaying relation between X-ray flux and VLF signal amplitude perturbation both during the minima and maxima phases of the solar cycle.  

 There has been no comprehensive and extensive study on the behaviour of the lower ionosphere over a solar cycle which explains the variations in lower ionospheric conditions, particularly the electron density distribution throughout an entire solar cycle. To achieve this, a long-term remote sensing study, such as, VLF propagation is necessary. We aim to undertake such a study. As a preliminary step, we conduct a straightforward numerical calculation to understand this behaviour.
 In this work using in-situ numerical simulation, we investigate the effects of the solar EUV (Lyman-$\alpha$ to be specific) and X-ray flux variation during C24 on the rate of ionization ($q$) in the D-region ionosphere. Hence, we estimate the long-term effect of solar activity cycle on D-region electron number density ($N_e$). We validate it with an observed long-term profile of a standard solar parameter, namely, the SPN and an ionospheric parameter, namely, the D-region electron density as estimated by the IRI-2020 model ($N_{e,iri}$). 

 The solar radiation flux has a direct correspondence with the Sunspot Number (SPN). This radiation, especially the UV radiation in the absence of any strong solar events (like, solar flares), directly affects the ionization rates in the lower ionosphere. The complex processes of electron density evolution, including various recombination, attachment and detachment processes are also involved. Hence, the investigation of the long-term variation of the lower ionosphere over an entire solar cycle in relation to the SPN is important. It provides a general understanding of how the lower ionosphere depends on solar conditions and paves the way for more detailed ionospheric observational studies.

 We opt for a well-accepted approach to numerically model the D-region electron number density ($N_e$) by solving the `electron continuity equation' \citep[][etc.]{whitten61, rowe70, anantha73,zigman07,basak13, palit16,palit18,nina18,chakraborty22a,chakraborty22c} in presence of the solar ionizing radiation. \citet{chakraborty20} first used this numerical model for the D-region ionosphere. Their model was capable of estimating the changes in the D-region electron density when the solar X-ray radiation activity is predominant. They used it on the D-region only in the presence of solar flare effects. Hence, their model can hardly be used when the solar EUV radiation is comparable to or overriding the solar X-ray energy while affecting the ionosphere. Whereas, our model is capable of handling the contribution from both solar EUV and X-ray radiations in D-region ionization, making it more suitable for both high and low solar activity phases.
 
 We look into the variation in solar Lyman $\alpha$ flux ($\phi_{L-\alpha}$), which is the primary contributor in ionizing D-region \citep{chubb57, nicolet60,kreplin62, yonezawa66, hinteregger81, rees89, bilitza2000, thomson01, basak13, chakraborty16, chakraborty20, chakraborty22b}, and that of solar X-ray flux ($\phi_{xr}$). We estimate the effects of $\phi_{L-\alpha}$ \& $\phi_{xr}$ in ionizing the D-region throughout C24. We present the altitude ($h$) and seasonal dependencies of the rate of ionization and consequently, the electron density during C24. We compute $N_e$ for $15^{th}$ day of every alternate month for the entire duration of C24 and observe the 'dual peak' nature during maxima which is also present in the observed profiles of SPN, $\phi_{L-\alpha}$ \& $\phi_{xr}$. We do a first-order validation of the outcomes with IRI-2020 modeled electron density ($N_{e,iri}$) values and conclude with possible explanations of the reported results. 

\section{Methodology}

We start with (i) the analysis of the solar Lyman-$\alpha$ ($\phi_\alpha$) and X-ray ($\phi_{xr}$) data recorded during C24. Then, (ii) we estimate the ionization rates ($q$) numerically using $\phi_\alpha$ and $\phi_{xr}$. (iii) Using $q$, we numerically compute the D-region electron density ($N_e$) profile during C24 using the solutions of `electron continuity equation'. (iv) Then, we extend the computations for $N_e$ across different altitudes of the D-region over two different geographical latitudes (e.g. 0$^\circ$ \& 45$^\circ$N). (v) Further, we consider SPN ($r$) during C24 and compare with the long-term variation of $N_e$. (vi) Finally, we take similar electron density values from the IRI-2020 model ($N_{e,iri}$) for necessary comparison with $N_e$. Here, we elaborately state the model and methodologies adopted at various mentioned stages of the work.

\subsection{Solar Extreme Ultraviolet (EUV) and X-ray flux ($\phi_{xr}$)}

We use the $15$-sec averaged $26$-$34$ nm solar flux for that period (coming from the entire solar disk) obtained from the Solar EUV Monitor (SEM) onboard the SOlar and Heliospheric Observatory (SOHO). The data is obtained from the Laboratory of Atmospheric and Space Physics (LASP), University of Colorado. Standard EUV spectrum ($\sim$ $5$-$130$ nm) from \cite{torr79} and \cite{torr85} is used in our calculation. This spectrum consists of photon counts not only in continuous range but in the discrete values also corresponding to various lines including the solar Lyman-$\alpha$ line ($121.56$ nm). The standard spectrum is normalized using the ratio of flux in the $26$-$34$ nm obtained from SOHO/SEM to that from the standard spectrum to obtain the spectra ($15$-sec average). These spectra are then used to find the rate of ionization throughout the day.

\begin{figure}
 \includegraphics[width=14.5cm,keepaspectratio]{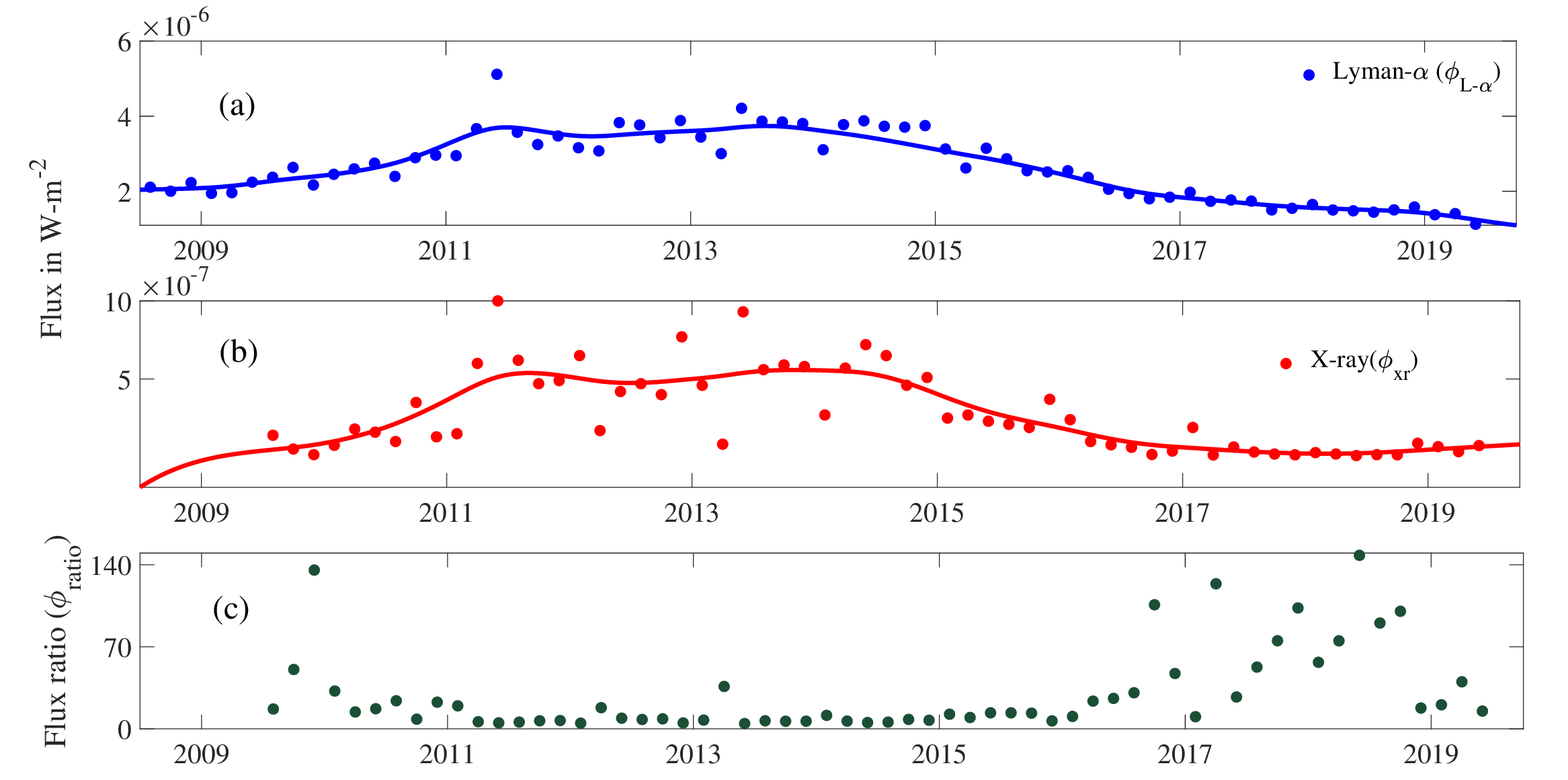}
 \caption{Variation of the daily average of (a) solar Lyman-$\alpha$ flux ($\phi_{L-\alpha}$), (b) solar X-ray flux ($\phi_{xr}$), and the (c) ratio of $\phi_{L-\alpha}$ and $\phi_{xr}$ ($\phi_{ratio}$) ($\phi_{ratio}$ is a dimensionless number) for the 15$^{th}$ day of every alternate month (i.e. Jan, Mar, May, Jul, Sep and Nov) during C24}
 \label{fluxes}
 \end{figure}

We take the GOES solar soft X-ray flux density ($\phi_{xr}$) in W-m$^{-2}$ unit from NCEI-NOAA. We consider only the background X-ray flux values for the corresponding days. As we are interested in exploring only the long-term effect and not any transient solar event, we ignore the effect of solar flares.
Besides, we compute the solar Lyman-$\alpha$ flux ($\phi_{L-\alpha}$) for the photons having wavelength $121.56$ nm. We estimate its daily average values for the 15$^{th}$ day of every alternate month during C24. We make a comparative estimation of the presence of Lyman-$\alpha$ flux and the soft X-ray flux for the whole period of C24. For this, we compute the ratio of $\phi_{L-\alpha}$ and $\phi_{xr}$ (Fig. \ref{fluxes}). 

\cite{venka23} also considered the Lyman-$\alpha$ flux for the entire span of C24. They showed the 'dual peak' in $\phi_{L-\alpha}$ profile, which is to some extent visible in our case. The 'dual peak' is also visible in $\phi_{xr}$ profile as well at the same time (Fig. \ref{fluxes}b). 

\subsection {Rate of ionization due to EUV radiation ($q_\alpha$)}

 Following Chapman's prescription as discussed in \cite{chakraborty16}, we use a methodology to compute $q_\alpha$ during different phases of solar activity. Chapman's formula (\cite{chapman31}) is used to compute $q_\alpha$ as a function of time ($t$) and altitude ($h$). The standard expression for $q_\alpha$ is given by,
 \begin{linenomath*}
 \begin{equation}  \label{eq1}
q_\alpha(h,t) = \sum_{i}^{} \int_{}^{} I_0(\nu,t)exp[-\sum_{j}^{} \sigma_j(\nu) \int_{h}^{\infty} n_jC_h(h,\chi)dh] \times \eta_i(\nu)\sigma_i(\nu)n_i(h)d\nu, 
 \end{equation}
\end{linenomath*}
where, $I_0(\nu, t)d\nu$ is the solar flux at the top of the atmosphere within the frequency range $\nu$ to $\nu + d\nu$. $\sigma_i(\nu)$ is the absorption cross section for the $i^{th}$ neutral component of air, which is a function of the energy of the photons. $n_i(h)$ is the concentration and $\eta_i(\nu)$ is the photo-ionization efficiency for the $i^{th}$ component. $C_h(h,\chi)$ is the `grazing incidence function' and it is discussed in \cite{rees89}. $\chi$ is the `solar zenith angle'. We adopt the standard profiles of $\sigma_i(\nu)$, $n_i(h)$ and $\eta_i(\nu)$, especially for D-region from \cite{torr79} and \cite{torr85}.  

\begin{figure}
\includegraphics[width=14.5cm,keepaspectratio]{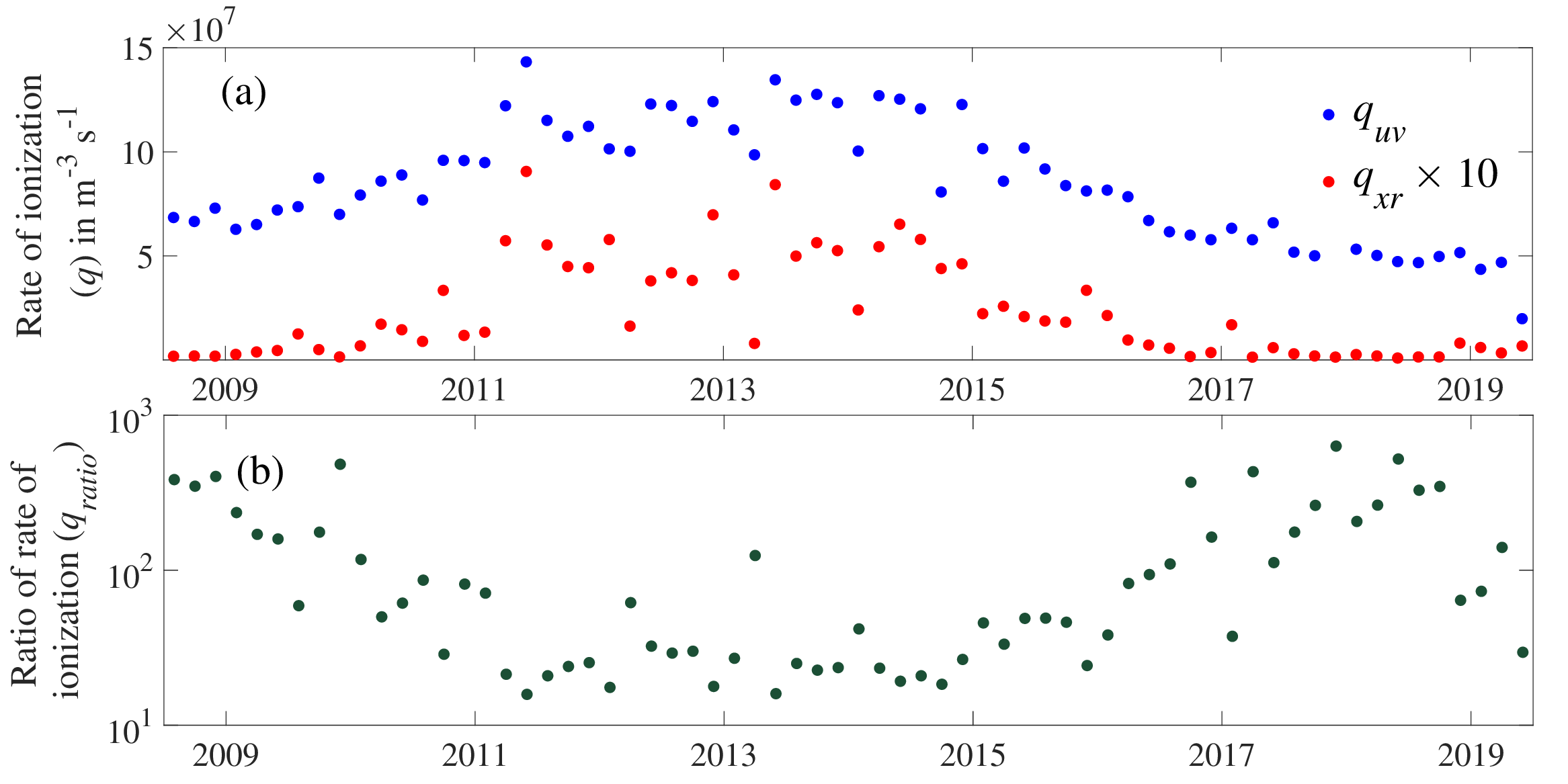}
\caption{Variations of (a) the rate of ionization due to solar EUV ($q_{uv}$) (blue dots) and X-ray ($q_{xr}$) (red dots) flux are plotted. (b) The ratio of $q_{uv}$ and $q_{xr}$ ($q_{ratio}$) during C24 over equatorial D-region at $80$ km altitude is shown.}
\label{qs}
\end{figure}

\subsection {Rate of ionization due to solar X-ray radiation ($q_{xr}$)}

The X-ray always plays a crucial role in ionizing the D-region ionosphere and becomes significant under high solar activity conditions. The complex chemistry behind the ionization mechanism is popularly simulated under justified assumptions using the GEANT4 Monte Carlo simulation technique (\cite{glukhov92}, \cite{haldoupis09}, \cite{palit13}, \cite{palit14gass}). Since, we deal mostly with ambient ionospheric conditions for estimating the rate of ionization ($q_{xr}$) as a function of $h$, we take a straightforward approach to compute $q_{xr}$ with the methods described in \cite{ratcliffe72}, \cite{budden88}, \cite{zigman07}, \cite{basak13}, \cite{chakraborty20} etc. \cite{chakraborty22c} established $h$ profile of $q_{xr}$ by incorporating the altitude profile of scale height ($H$) as mentioned in \cite{mitra52}, \cite{basak13}, \cite{chakraborty20} etc. To get $h$ profile of $q_{xr}$, we follow a similar methodology. The expression of $q_{xr}$ is given by,
\begin{linenomath*}
\begin{equation}  \label{eq2}
q_{xr}(h,t) = \frac{\phi_{xr}(t)}{exp(1)H(h) \rho} cos(\chi),  
\end{equation}
\end{linenomath*}
where, $\rho$ ($=34 eV$) is the amount of energy required to create an electron-ion pair. The profile of $\chi$ is taken from \cite{rees89}.
Now, adding Eq. \ref{eq1} and Eq. \ref{eq2}, we get the total rate of ionization both due to EUV and X-ray. It is as follows,
\begin{linenomath*}
\begin{equation}  \label{eq3}
q(h,t) = q_{uv}(h,t) + q_{xr}(h,t).
\end{equation}
`\end{linenomath*}

\subsection{Computation of D-region electron density profile ($N_e(h,t)$)}

The D-region `electron continuity equation' is one of the simple tools to estimate the electron density profile ($N_e$) by suitably taking into account the ionization and recombination processes occurring in that region (\cite{whitten61}, \cite{rowe70}, \cite{anantha73}, \cite{zigman07}, \cite{basak13}, \cite{palit16}, \cite{palit18}, \cite{nina18}, \cite{chakraborty20}, \cite{chakraborty22a}, \cite{chakraborty22c} etc.). Here, we take an approximated form of this equation, 
 \begin{linenomath*}
 \begin{equation}  \label{eq4}
\frac{dN_e(h,t) }{dt} = \frac{q(h,t)}{1 + \lambda(h)} - \alpha_{eff}(h) N_e^2(h,t),
 \end{equation}
\end{linenomath*}
where, $N_e(h,t)$ is the altitude profile of D-region electron density at different instants of time, $\lambda(h)$ is the ratio of negative ion to electron number densities and $\alpha_{eff}(h)$ is the `effective recombination coefficient', which effectively takes into account all kinds of recombination processes in respective proportions. We adopt the required profiles of $\lambda(h)$ and $\alpha_{eff}(h)$ from \cite{palit15} and \cite{chakraborty22c}. We take $q(h,t)$ from Eq. (\ref{eq3}) and perform the standard method as described in \cite{chakraborty20} to solve Eq. (\ref{eq4}) to obtain $N_e(h,t)$. We repeat it for the 15$^{th}$ day of every alternate month (i.e. Jan, Mar, May, Jul, Sep and Nov) within the span of C24. \cite{chakraborty20} and \cite{chakraborty22c} reported a noteworthy latitude-longitude dependence in $N_e(h,t)$ profile. We choose to analyse the D-region ionosphere over $88^{\circ}$E longitude and $45^{\circ}$N latitude (E88, N45) at $12$ noon local time. Then, we repeat the same at the equator ($0^{\circ}$N longitude) while computing the altitude profile of $N_e(h)$. We compare our numerical results regarding $N_e(h,t)$ with its IRI-2020 counterpart ($N_{e,iri})$.

\begin{figure}
\includegraphics[width=14.5cm,keepaspectratio]{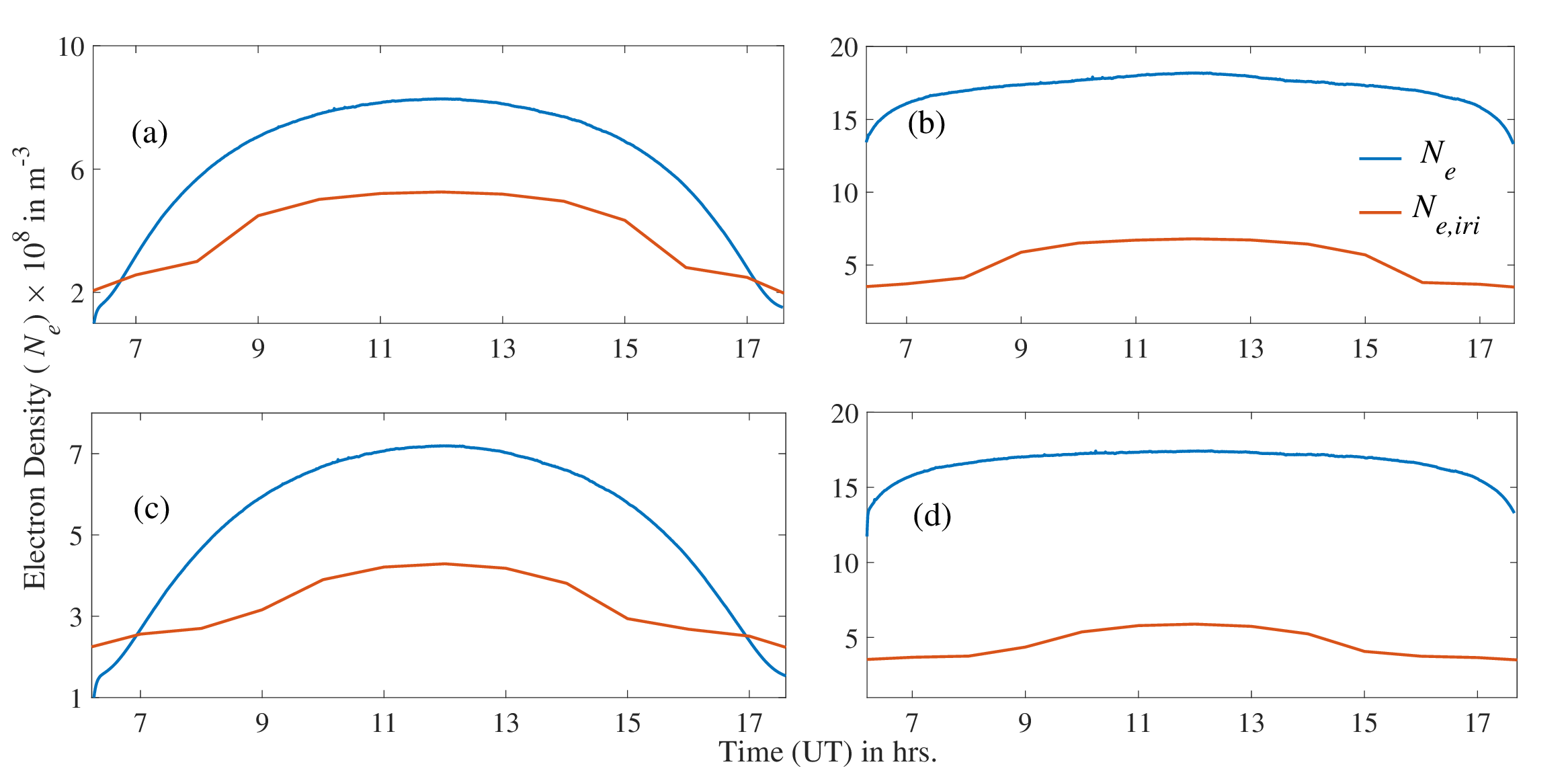}
\caption{Daily variations of electron densities estimated from our model and IRI-2020 over the equator (a \& b) and N$45$ (c \& d) at two different altitudes, namely, $75$ km (a \& c) and  80 km (b \& d) are shown.}
\label{diur}
\end{figure}

To validate our electron density calculation model and to compare its outcome with IRI-2020, we calculate the daily variation of electron density at the two latitudes, namely, at the equator and N$45$. In Fig. \ref{diur}, we include the electron density variation at two different D-region altitudes ($h$), i.e., $75$ km and $80$ km.

\subsection{Sunspot Number (SPN)}

We take the relative SPN ($r$) from the standard concept of `Wolf's Sunspot Number Formula' (\cite{wolf1851}, \cite{wolf1856}). It is stated as,
 \begin{linenomath*}
 \begin{equation}  \label{eq5}
r = k(10g+n),
 \end{equation}
\end{linenomath*}
where, `$k$' is a correction factor. The `$g$' is the number of identified sunspot groups and `$n$' is the number of individual sunspots. We take the monthly average of $r$ ($R$) for every alternate month for the entire duration of C24. We compare the long-term variation of $N_e$ with the similar variation of $R$. We perform a standard polynomial fitting to both $N_e$ and $R$ to understand the gross nature of variation (Fig. \ref{ner}c).

\section{Results \& Discussions}

We model the long-term variability of D-region electron density ($N_e$) during the solar cycle-24 (C24). We consider the D-region ionization rates, $q_{uv}$ and $q_{xr}$ during C24 due to two prominent ionizing agents, namely, the solar-UV flux ($\phi_{uv}$) and solar X-ray flux ($\phi_{xr}$). Then, we numerically solve the `electron continuity equation' (Eq. \ref{eq4}) to obtain the altitude profile of the D-region electron density ($N_e$) at $12$ noon local time on 15$^{th}$ day of every alternate month of C24. We bring in the average sunspot number ($R$) and the IRI-2020 model-generated electron density ($N_{e,iri}$) to compare with our results. 
  
During C24, both $q_{xr}$ and $q_{uv}$ have significant variations following the overall solar activity (Fig. \ref{qs}a). \cite{reid76} reported a similar increment in ionization during the solar maxima. Especially during the solar minima (which is at the beginning and end of a solar cycle), we note that $q_{xr}$ values are significantly smaller in comparison to $q_{uv}$. In Fig. \ref{qs}a, we multiply $q_{xr}$ with a factor of `$10$' just to put it in a comparable scale to plot it together with $q_{uv}$. $q_{ratio}$ is shown in Fig. \ref{qs}b, which is found to range between $10$ and $150$. Interestingly, we report that $q_{xr}$ \& $q_{uv}$ have roughly comparable values during solar maxima and $q_{ratio}$ values are nearly within $10$ and $30$. On the other hand, $q_{ratio}$ goes beyond $100$ during solar minima. It infers that ambient UV radiation is much more dominant than ambient X-ray radiation in ionizing the D-region.  

We find a prominent extra peak in the cases of both the profiles of $\phi_{L-\alpha}$ and $\phi_{xr}$ during November 2011 (Fig. \ref{fluxes}). The cause of this is the comparatively higher activity of the sun for a shorter period of time. It is the `Gnevyshev peak' (\cite{gnevyshev63}). It has resulted is respective enhancements in the values of $q_{uv}$ and $q_{xr}$, i.e. the effect of the `Gnevyshev peak' on the D-region ionosphere in evident (Fig. \ref{qs}). The correlation coefficient value between Figs. \ref{fluxes}a and \ref{qs}(a) (blue dots) is $0.96$ and same between Figs. \ref{fluxes}b and \ref{qs}(a) (red dots) is $0.99$.  

\subsection{Seasonal dependence of the altitude profile of $q$ (\& $N_e$) during C24}

$q$ (i.e., $q_{uv} + q_{xr}$), and subsequently $N_e$ have notable seasonal dependencies due to variation in solar zenith angle ($\chi$). \cite{chakraborty20} reported a seasonal dependency of $N_e$. They claimed that the gross $N_e$ values are larger during summer than winter in northern hemispheric regions. As we perform numerical simulation of in-situ solar radiation effect throughout C24 starting from solar ionization calculation by UV and X-ray, we delve deeper into the understanding of the ionization characteristics and its altitude, latitude, as well as, solar cycle variations.  

\begin{figure}
 \includegraphics[width=14.5cm,keepaspectratio]{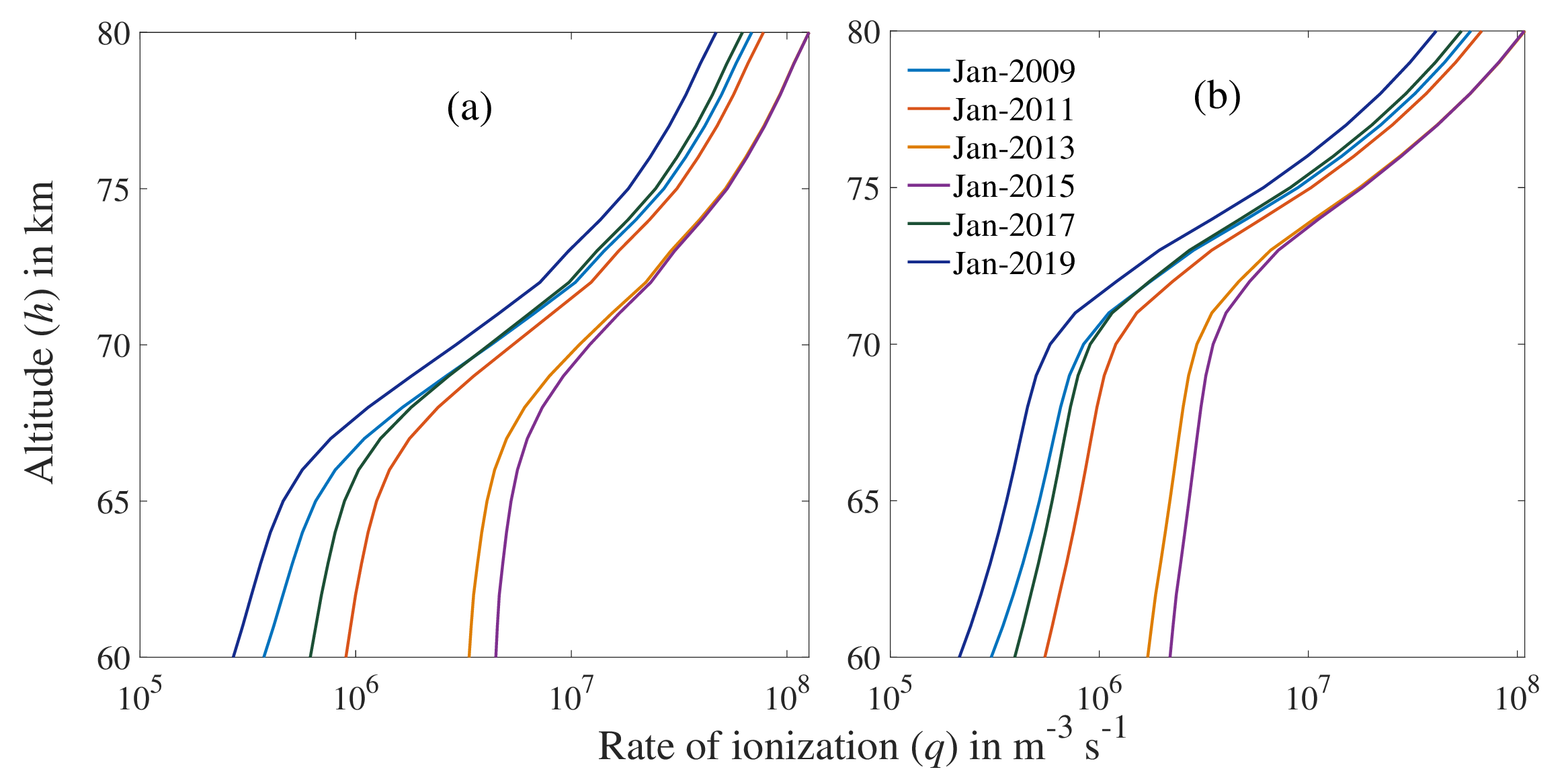}
 \caption{Altitude ($h$) profiles of the rate of total ionization ($q$) at D-region for the month of January of every alternate year during C24 over (a) equator and (b) N45.}
 \label{qh}
 \end{figure}

To find the variation in $h$ dependency of $q$, we compute $h$ profiles of $q$ at $12$ noon of every 15$^{th}$ day of January of every alternate year during C24. We repeat it for the equatorial latitude (Fig. \ref{qh}a) and for the N45 latitude (Fig. \ref{qh}b). We note the following from the analysis.
 
 (i) $h$ dependency of $q$ has greater prominence at higher altitudes. The variation in chemical properties of successive lower ionospheric layers is one of the reasons this. The fact that the differences in the ratio of $\chi$'s between two successive pairs of $h$'s are not the same at all altitudes. It is also a possible cause for such variation of $q$ (\cite{rees89}). Also, the relative ionization is lesser in lower altitudes. This is because solar radiation passes through the ionosphere by experiencing successive absorption. Hence, as the radiation travels closer to the earth's surface, it becomes gradually less effective in terms of ionizing the atmosphere.  

 (ii) We report significant solar cycle dependency in $q$-$h$ profiles. The total rate of ionization ($q$) decreases gradually as solar radiation travels down to $\sim$ $80$ km and thereafter decreases rapidly. It creates a `hinge' like structure in general and interestingly over both the latitudes (Fig. \ref{qh}). For equatorial latitude, this `hinge' appears at around $h$=$67$ km and for N45 latitude it appears at $h$=$70$ km. All these $q$'s are plotted for January when solar radiation was relatively slant on N45 latitude ($\approx60^{\circ}$) compared to the same in the equatorial latitude ($\approx15^{\circ}$). The difference in the inclination between the two latitudes could be responsible for such different heights of the hinges in the ionization vs. height ($q$-$h$) profile (Figs. \ref{qh}a \& b).

 (iii) The $q$-$h$ profiles of different years are different from each other with the highest values of rate of ionization during January 2013 and the lowest during January 2019. Furthermore, the $q$-$h$ profiles of January 2019 (light-blue curve) show lesser values than that of January 2009 (blue curve). It implies that the solar activity has been dimmer in terms of overall radiation level. Hence, the ionization of the lower ionosphere during the second minima of the C24 around 2019 was the lowest (Fig. \ref{qh}).

\begin{figure}
\includegraphics[width=14.5cm,keepaspectratio]{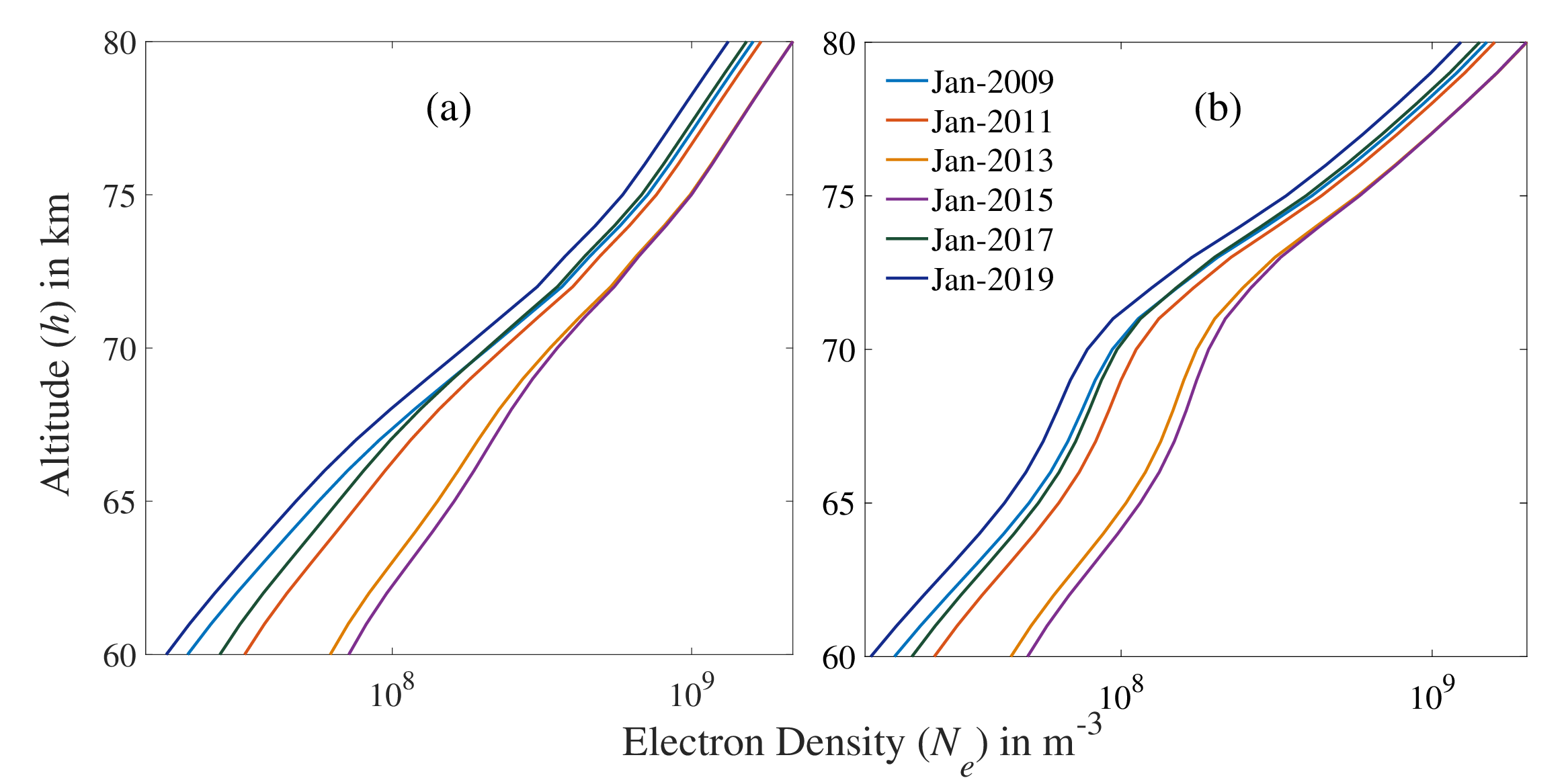}
\caption{Altitude ($h$) profiles of the D-region electron density ($N_e$) for the month of January of every alternate year during C24 over (a) equator and (b) N45 as obtained from the computations using the rates of total ionization ($q$) shown in Fig. \ref{qh}}
\label{neh}
\end{figure}
\begin{figure}
\includegraphics[width=14.5cm,keepaspectratio]{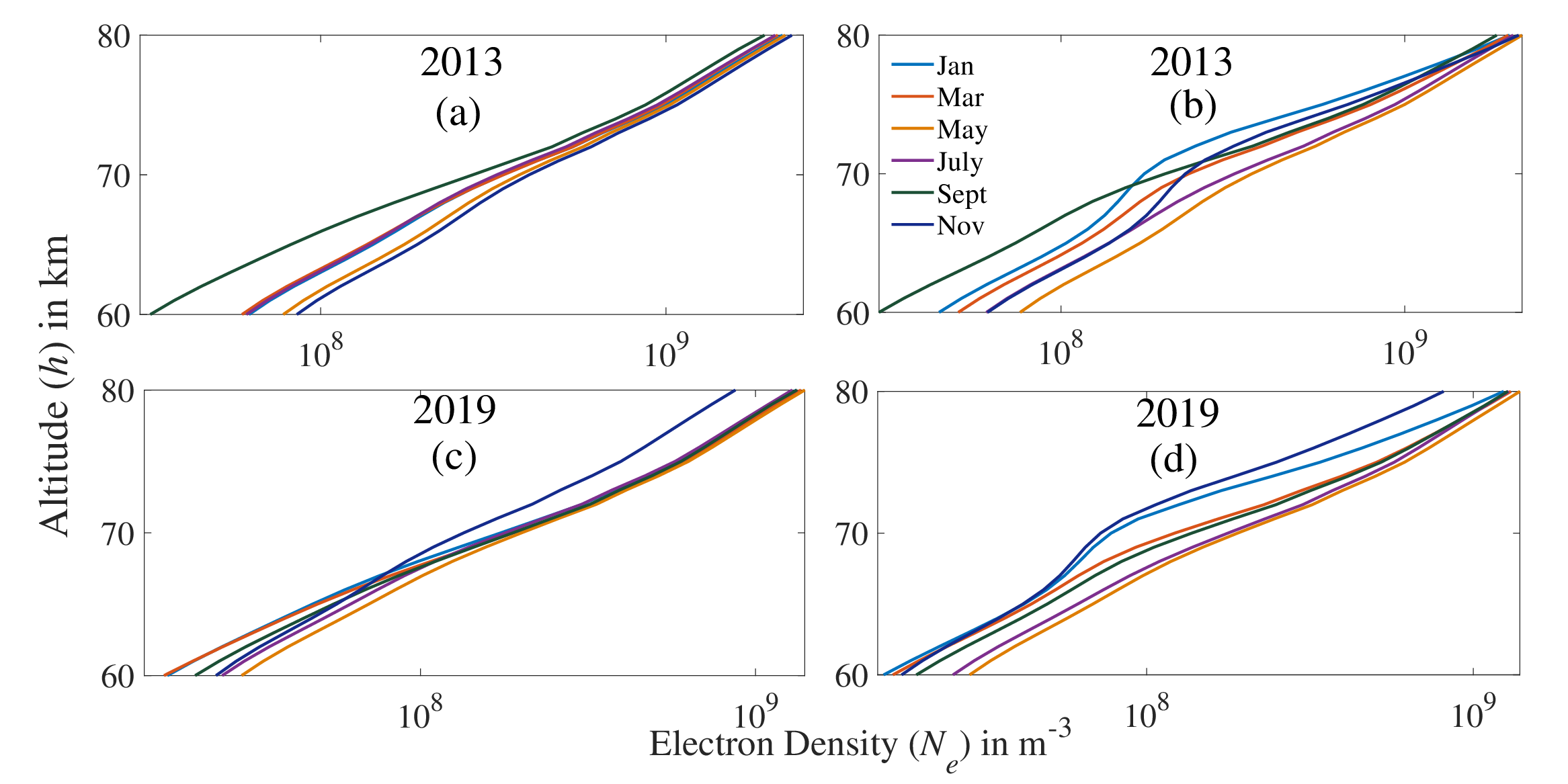}
\caption{Altitude ($h$) profiles of D-region electron density ($N_e$) for the months of Jan, Mar, May, Jul, Sep \& Nov of 2013 and 2019 over (a \& c) equator and (b \& d) N45 are shown respectively.}
\label{nehs}
\end{figure}

All such dependence of $q$ over the solar cycle and latitude has been manifested in the $N_e$-$h$ profile as presented in Fig. \ref{neh}. We present here only the electron density altitude profiles at $12$ noon local time instead of a diurnal variation. We have found the variation of electron density at $80$ km from  $\sim 1.2 \times 10^9$  to  $\sim 2.3 \times 10^9$ m$^{-3}$ from minima (2019) to maximum (2013), which is almost twice, whereas, at $60$ km. it is from $\sim 2 \times 10^8$ to  $\sim 8 \times 10^8$ m$^{-3}$, i.e., nearly $4$ times. Subsequently, we check the seasonal dependency of the $N_e$-$h$ profile. For this, we choose a particular year during solar minima, namely, 2019, and one during maxima, namely, 2013. We compute the $N_e$-$h$ for the 15$^{th}$ day of every alternate month of those two years. We show $N_e$-$h$ profiles for 2013 for equatorial latitude (Fig. \ref{nehs}a) \& N45 latitude (Fig. \ref{nehs}b). We repeat the procedure for 2019 (Fig. \ref{nehs}c \& d). We note the following.

(i) $N_e$-$h$ profile has a significant seasonal dependency in both the mentioned latitudes. There are clear spreads in $N_e$ values as one goes from January to November, though the spread is not uniform throughout the altitude range. For example, the $N_e$-$h$ profile in equatorial latitude in the year 2013 (Fig. \ref{nehs}a) shows that $N_e$ of November is the lowest for $h>70$ km, but for $h<70$ km, $N_e$ of July (violet curve) has the minimum value. (ii) Seasonal variation of $N_e$-$h$ profiles has a notable latitude dependency. The curves are more spread for N45 latitude \ref{nehs}(b \& d) during both solar maxima and minima. The slant nature of solar radiation over N45 could be a possible reason for such variation. (iii) Seasonal variation of $N_e$-$h$ profiles has a noteworthy dependency on the solar cycle as well. During solar minima, smaller values of $q$'s at all the heights lead to smaller values of $N_e$'s. Also, we see more spread in $N_e$ values in both the latitudes \ref{nehs}(c \& d).

\subsection{Long-term variation of $N_e$ and average SPN ($R$)}

\begin{figure}
 \includegraphics[width=14.5cm,keepaspectratio]{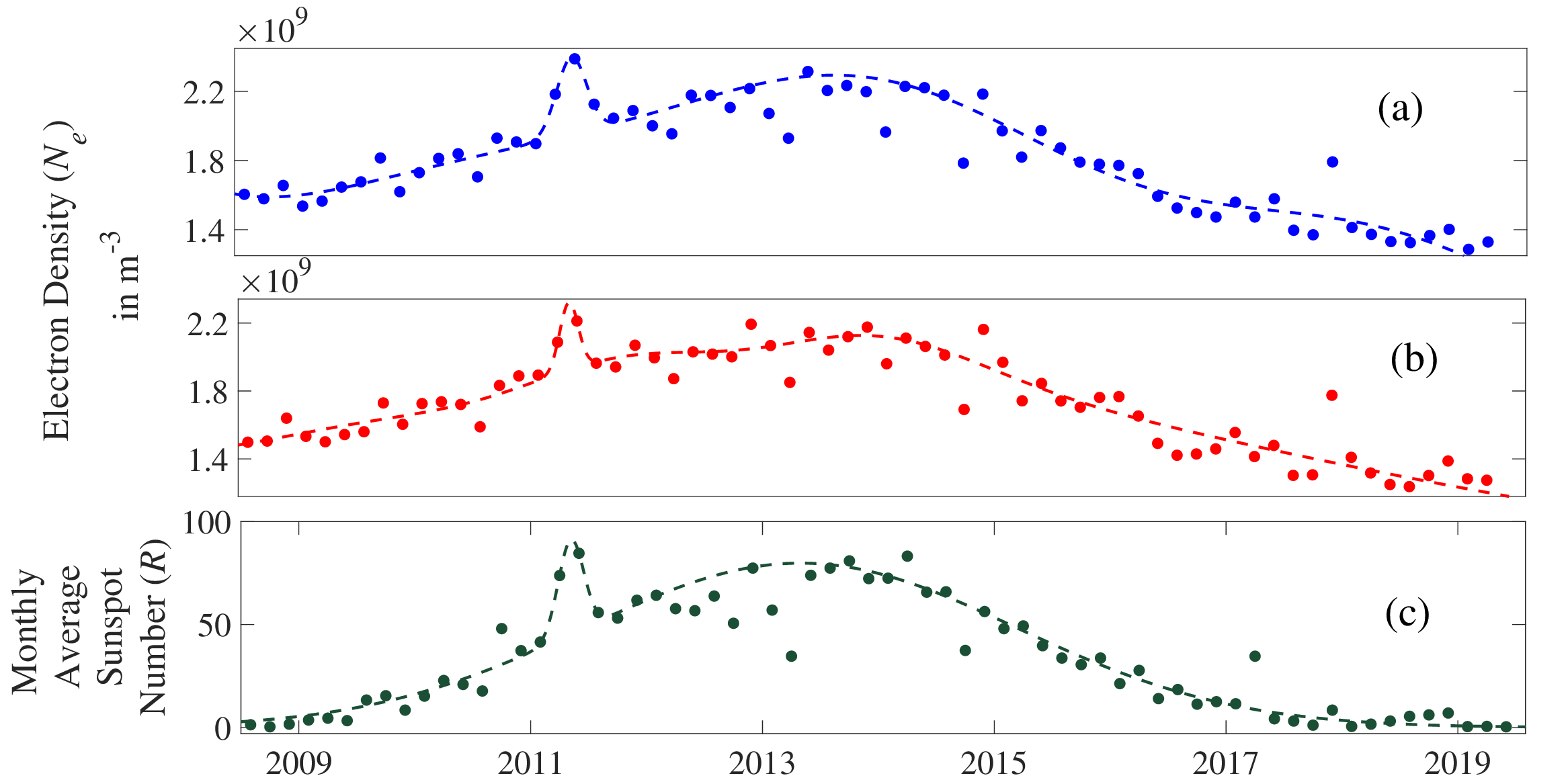}
 \caption{Long-term variation of D-region electron density ($N_e$) (at $80$ km altitude) is shown over (a) the equator and (b) N45 during C24. (c) A similar variation of the monthly average of SPN ($R$) is shown. A standard polynomial fitting is done.}
 \label{ner}
 \end{figure}

 We compute $N_e$ for $12$ noon local time at $h$=$80$ km for 15$^{th}$ day of every alternate month during C24. We repeat it for both equatorial (Fig. \ref{ner}a) and N45 latitudes (Fig. \ref{ner}b). Besides, we plot the monthly averaged SPN ($R$) for a comparative analysis (Fig. \ref{ner}c) as the SPN is one of the most precise and handy indicators of the solar cycle (\cite{newkirk82}). We do polynomial fittings for a better understanding of the overall nature of variations of $N_e$ and $R$.

 (i) Evidently, there is a significant variation of $N_e$ across the entire span of C24 which depends on latitude as well. 
 (ii) The `dual peak' nature appears in $N_e$, as well as in $R$ profile. The first peak appeared from September to November 2011 and the second broader peak appeared from December 2013 to May 2014 both in $N_e$ (over both the latitudes) and $R$ profiles. In equatorial latitude (Fig. \ref{ner}a), $N_e$ starts increasing from $\sim 1.8\times10^9$ m$^{-3}$ during the solar minima (2009) and reaches the 1st maxima ($\sim 2.3\times10^9$ m$^{-3}$) during November 2011. Thereafter, it decreases to $\sim 2\times10^9$ m$^{-3}$ from January 2012 to September 2013, and again touches the 2nd maxima ($\sim 2.2\times10^9$ m$^{-3}$) during March 2014. It decays finally to a long-duration minima phase. A similar `dual peak' is reported in $N_e$ profile over N45 also, but this time the respective $N_e$ values are different from those over the equatorial latitude for obvious reasons (\cite{chakraborty20}). Since, $N_e$ is directly affected by solar activity level, it follows the `dual peak' as shown in $R$ and $\phi$ (Fig. \ref{fluxes}) profiles as well. (iii) Although, the change in $R$ from solar minima to maxima is $\sim 80$-$90\%$, and the same in $N_e$ is found to be $\sim 70\%$ over equatorial latitude and $\sim 55\%$ over N45.

\subsection{Collocation of $N_e$ and $N_{e,iri}$ with $R$}

\begin{figure}
\includegraphics[width=14.5cm,keepaspectratio]{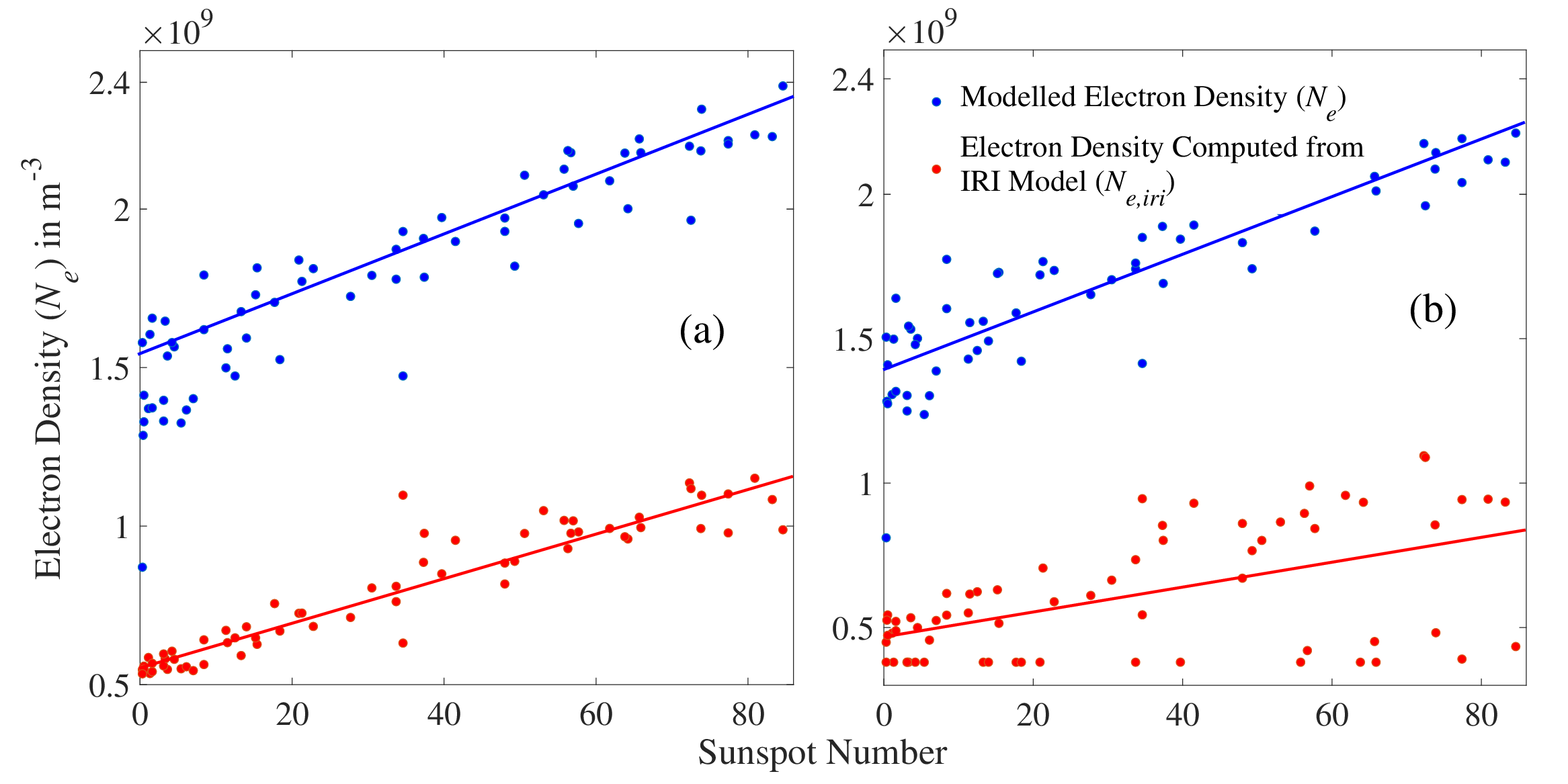}
\caption{Variation of the modeled D-region electron density ($N_e$) (at $80$ km altitude) (blue) and its IRI-2020 counterpart ($N_{e,iri}$) (red) during C24 over (a) the equator and (b) N45 is plotted as a function of the monthly average of SPN ($R$). A linear fitting is done on them.}
\label{ner_line}
\end{figure}

We plot the computed $N_e$ and IRI-2020 modeled electron density profiles ($N_{e,iri}$) at $h$=$80$ km as a function of $R$ during C24 for both the equatorial (Fig. \ref{ner_line}a) and N45 (Fig. \ref{ner_line}(b)) latitudes. We report the following observations.

(i) Both $N_e$ \& $N_{e,iri}$ vary almost linearly with $R$. As the increase in $R$ signifies an enhancement in solar activity level (\cite{hoyt98}, \cite{clilverd06}, \cite{lukianova11}, \cite{clette14}), we note that the electron density in the D-region increases linearly with increasing solar activity level. \cite{bullen61} reported that the ionospheric F-region electron density profile follows a nearly similar tendency with a running average of $R$. 

(ii) We report that $N_e$ values are nearly thrice than $N_{e,iri}$, and the gradient of slopes ($s$) also differ for $N_{e,iri}$ over both the latitudes. The values of gradient of slope ($s$) and goodness of fit ($\chi^2$) of $N_e$ (\& $N_{e,iri}$) versus $R$ plots (Table 1). We report a greater slope for both $N_e$ and $N_{e,iri}$ profiles at the equatorial latitude ($s=1.016\times10^7$ \&  $7.023\times10^6$ respectively) than N45 ($s=9.9\times10^6$ \& $4.3\times10^6$ respectively). Both $s$ and $\chi^2$ signify the quantitative relation between those parameters. The IRI-2020 model generally provides more accurate electron density values at higher ionospheric regions, such as the F-region. However, studies by \cite{danilov95}, \cite{bilitza98}, \cite{coisson06}, \cite{singh21} and others have reported that the IRI model tends to underestimate electron density values at lower ionospheric altitudes, like the D-region. Additionally, the lower ionospheric models face limitations due to their heavy reliance on available chemical rate parameters, which are mainly derived from empirical studies due to the lack of observational capabilities in these regions. The chemical reaction parameters, and consequently the effective recombination coefficients calculated from them, are not precisely known and can vary within a range (see \cite{glukhov92}, \cite{palit13} \cite{palit15}). Therefore, we are not particularly inclined to compare our results with other models like IRI in terms of exact electron density values. Instead, we focus on the relative variations in electron density with respect to positional, temporal parameters, and solar conditions. 

(iii) Lastly, we report that $N_{e,iri}$ of N45 latitude (Fig. \ref{ner_line}b) shows more scattered points in the plot than that for the equatorial region. 

\begin{table}
\caption{Gradient of slopes ($s$) and goodness of fit ($\chi^2$) of the linear fitting on $N_e$-$R$ \& $N_{e,iri}$-$R$ profiles in Fig. \ref{ner_line}}
\centering
\begin{tabular}{|c|c|c|c|}
\hline
Latitude  & Electron density & Gradient of slope ($s$) in $m^{-3}$  & Goodness of fit ($\chi^2)$\\
\hline
   $equator$  & $N_e$  &  $1.016\times10^7$  &  0.8195  \\
   $equator$  &  $N_{e,iri}$  &  $7.023\times10^6$  &  0.8913  \\
   $45^{\circ}N$  &  $N_e$  &  $9.9\times10^6$  &  0.8003  \\
   $45^{\circ}N$  &  $N_{e,iri}$  &    $4.3\times10^6$  &  0.2881  \\
\hline
%\multicolumn{2}{l}{$^{a}$Footnote text here.}
\end{tabular}
\end{table}

\section{Conclusion}
The Earth's ionosphere, a partially ionized layer shaped by solar radiation, naturally responds to changes in solar activity. While long-term solar effects on the upper ionosphere are well-studied, similar research on the lower D-region is limited, primarily focusing on short-term events, like solar flares. Due to the challenges of directly investigating the lower ionosphere, remote sensing techniques using VLF radio waves are commonly employed. This study utilizes computational modeling to examine the long-term modulation of the D-region during a complete solar cycle. We leverage extensive data on solar radiation and the chemical composition of lower ionospheric layers to create a comprehensive picture. Additionally, we account for seasonal and latitudinal variations, recognizing that the angle of solar radiation affects different areas of earth's ionosphere throughout the year and can significantly impact the properties of the lower ionosphere. We also compare our findings with the same using an established model of the ionosphere, namely the IRI-2020 model and corroborate the study with the most reliable solar activity indicator, the SPN.  

We have got a quantitative estimation of how the two most prominent ionizing components of solar radiation, namely the EUV and soft X-ray compare to each other in terms of modulating the lower ionosphere. We have computed both the ionization rates, namely, $q_{uv}$ and $q_{xr}$. We see that the contribution of the soft X-ray in ionizing the part of the ionosphere compared to that of the EUV is comparatively smaller during solar minima. But, during solar maxima, ionization due to soft X-ray dominates. The manifestation of the double peak during solar maxima (called the {\it{Gnevyshev gaps}}) in terms of electron density has also been included in the results. We have found a linear correlation between the SPN and the average electron density throughout the cycle. Though the absolute values of the electron densities ($N_e$) from our numerical model and those from the IRI-2020 model ($N_{e,iri}$) are not close, they show the same correlation with the SPN ($R$). 

We are in the process of investigating the long-term effect of solar cycle modulation on the D-region ionosphere with VLF observation and will put forward a comparative overview between the numerical and observational outcomes in some later studies.

\acknowledgments
Authors thank NCEI-NOAA for solar X-ray and sunspot number data, Laboratory of Atmospheric and Space Physics (LASP), University of Colorado for EUV light-curve, and International Reference Ionosphere model for electron density data. Sayak Chakraborty acknowledges the support of DST-INSPIRE fellowship (IF200266), Department of Science and Technology, Govt. of India.

\section*{Open Research Section}
The 15-sec average $26$-$34$ nm solar flux data for that period is taken from the Solar EUV Monitor (SEM) onboard the SOlar and Heliospheric Observatory (SOHO), Laboratory of Atmospheric and Space Physics (LASP), University of Colorado ({\it {https://lasp.colorado.edu/}}). Standard EUV spectrum ($\sim$ $5$-$130$ nm) is taken from \cite{torr79} and \cite{torr85}. The GOES solar soft X-ray flux density is taken from NCEI-NOAA ({\it {https://www.ncei.noaa.gov/data/}}).

%%%%%%%%%%%%%%%%%%%%%%%%%%%%%%%%%%%%%%%%%%%%%%%
% REFERENCES and BIBLIOGRAPHY
%
 \bibliography{ref} %don't specify the file extension
% don't specify bibliographystyle
%
%%%%%%%%%%%%%%%%%%%%%%%%%%%%%%%%%%%%%%%%%%%%%%%

%\bibliography{ enter your bibtex bibliography filename here }

%Reference citation instructions and examples:
%
% Please use ONLY \cite and \citeA for reference citations.
% \cite for parenthetical references
% ...as shown in recent studies (Simpson et al., 2019)
% \citeA for in-text citations
% ...Simpson et al. (2019) have shown...
%
%
%...as shown by \citeA{jskilby}.
%...as shown by \citeA{lewin76}, \citeA{carson86}, \citeA{bartoldy02}, and \citeA{rinaldi03}.
%...has been shown \cite{jskilbye}.
%...has been shown \cite{lewin76,carson86,bartoldy02,rinaldi03}.
%... \cite <i.e.>[]{lewin76,carson86,bartoldy02,rinaldi03}.
%...has been shown by \cite <e.g.,>[and others]{lewin76}.
%
% apacite uses < > for prenotes and [ ] for postnotes
% DO NOT use other cite commands (e.g., \citet, \citep, \citeyear, \nocite, \citealp, etc.).
%

\end{document}